\documentclass[12pt]{article}
\usepackage{epsfig}
\title{Condensates in Quantum Chromodynamics}
\author{B.L.Ioffe\\
Institute of Theoretical and
Experimental Physics,\\
B.Cheremushkinskaya 25, 117218 Moscow,Russia}
\date{}
\begin{document}
\maketitle

\newcommand{\be}{\begin{equation}}
\newcommand{\ee}{\end{equation}}

\def\la{\mathrel{\mathpalette\fun <}}
\def\ga{\mathrel{\mathpalette\fun >}}
\def\fun#1#2{\lower3.6pt\vbox{\baselineskip0pt\lineskip.9pt
\ialign{$\mathsurround=0pt#1\hfil##\hfil$\crcr#2\crcr\sim\crcr}}}

\begin{center}
{\bf \large Abstract}
\end{center}

\vspace{5mm}

The paper presents the short review of our to-day knowledge of
vacuum condensates in QCD. The condensates are defined as vacuum
averages of the operators which arise due to nonperturbative
effects. The important role of condensates in determination of
physical properties of hadrons and of their low-energy
interactions in QCD is underlined. The special value of quark
condensate, connected with the existence of baryon masses is
mentioned. Vacuum condensates induced by external fields are
discussed. QCD at low energy is checked on the basis of the data
on hadronic $\tau$-decay. In the theoretical analysis the terms of
perturbation theory (PT) up to $\alpha^3_s$ are accounted, in the
operator product expansion (OPE) - those up to dimension 8. The
total probability of the decay $\tau \to hadrons$ (with zero
strangeness) and of the $\tau$-decay structure functions are best
described at $\alpha_s(m^2_{\tau}) = 0.330 \pm 0.025$. It is shown
that the Borel sum rules for $\tau$-decay structure functions
along the rays in the $q^2$-complex plane are in agreement with
the experiment with the accuracy $\sim 2\%$ at the values of the
Borel parameter $\vert M^2 \vert > 0.8 GeV^2$. The magnitudes of
dimensions 6 and 8 condensates were found  and the limitations on
gluonic condensates were obtained. The sum rules for the charmed
quark vector currents polarization operator was analysed in 3
loops (i.e., in order $\alpha^2_s$). The value of charmed quark
mass (in $\overline{MS}$ regularization scheme) was found to be:
$\overline{m}_c(\overline{m}_c) = 1.275 \pm 0.015 GeV$ and the
value of gluonic condensate was estimated: $\langle 0 \vert
(\alpha_s/\pi) G^2 \vert 0 \rangle = 0.009 \pm 0.007 GeV^4$. The
general conclusion is: QCD described by PT + OPE is in a good
agreement with experiment at $Q^2 \ga 1 GeV^2$.

\newpage

\begin{center}
{\bf \large 1. A few words about Igor' Vasilyevich.}
\end{center}


It is a great honour and at the same time a great pleasure for me
to write a paper to the issue of Physics of Atomic Nuclei
dedicated to 100-anniversary of Igor' Vasilyevich Kurchatov
birthday. Kurchatov was a very extraordinary person: an organizer
of the highest class, I know nobody with such excellent
organization abilities. Without him, the soviet atomic project
would not be, perhaps, realized, at the least in such short time.
Igor' Vasilyevich  had the strongest sense of responsibility not
only for the work entrusted to him -- the atomic project -- but
for the much wider -- for the fortune of the science in our
country and moreover, for the fortune of the whole mankind. I
shall give an episod not well known. As witnesses A.P.Aleksandrov
[1], Kurchatov was deeply depressed when coming back from the
tests of the first hydrogen bomb (those who were present at the
tests noticed the same). He said: "What a terrible thing we have
made. The only item we should bother about, is to forbid all of
this and to exclude nuclear war". In March 1954 Kurchatov,
Alikhanov, Kikoin and Vinogradov had written the paper, where they
concluded, that "-- the mankind is in front of the menace of the
end of all of the life on the Earth". The paper was also signed by
the Minister of the Middle Machine  Building V.A.Malyshev who had
sent it to Malenkov, Khrushchev and Molotov. Khrushchev, however,
had rejected the paper,  calling the words on a possible ruin of
the world civilization "theoretically wrong and politically
harmful" (see [2]). The position of the soviet leaders remained as
before: the world war should lead to the ruin of the capitalism.

The same responsibility was inherent to Igor' Vasilyevich when constructing
atomic reactors and atomic powerstations -- I gave examples of this earlier
[3]. I think that if Kurchatov would alive, RBMK reactors, principally  not
safe as physical systems, would not be constructed and we would avoid the
Chernobyl catastrophe. But on the other hand, Igor' Vasilyevich was a person
of his time ... (see [3]).

He liked the science and, in the first turn, his main
speciality-nuclear physics. He was deeply interested in the
development of the elementary particle physics and he thought that
it is necessary to develop such investigations in the USSR. He
supported the suggestion of Alikhanov and Vladimirsky of
construction at ITEP of the 7 GeV hard-focusing proton
accelerator, and then, using the ITEP project, of the 50 GeV
proton accelerator (later, 70 GeV) near Serpukhov. In 1954 such
decision had been adopted at the meeting of the chaired by
Kurchatov Scientific-Technical Council of the Middle
Machine-Building Ministry.

\vspace{7mm}

\centerline{\bf \large 2. Introduction}

\vspace{5mm}

 Nowadays, it is reliably established that the true
(microscopic) theory of strong interaction is quantum
chromodynamics (QCD), the gauge theory  of interacting quarks and
gluons. It is also established, that unlike, e.g., quantum
electrodynamics (QED), the vacuum in QCD has a nontrivial
structure: due to nonperturbative effects, i.e. not admitting the
expansion in the interaction constant (even if it is small) in QCD
vacuum persist non-zero fluctuations of gluonic and quark fields.
(Examples of such kind of nonperturbative fields are instantons
[4] -- classical solutions of equations for gluonic field, which
realize the minimum of actions in the QCD Lagrangian\footnote{This
statement refers to Euclidean space, in the Minkowsky space,
instantons realize tunneling transitions between Hilbert spaces
with different topological quanum numbers.}. The nontrivial
structure in QCD manifests itself in the presence of vacuum
condensates, analogous to those in the condensed matter physics
(for instance, spontaneous magnetization).  Vacuum condensates are
very important in elucidation of the QCD structure and in
description of hadron properties at low energies. Condensates, in
particular, quark and gluonic ones, were investigated starting
from 70-ties.  Here, first, it should be noted the QCD sum rule
method by Shifman, Vainshtein, and Zakharov  [5], which was based
on the idea of the leading role of condensates in the calculation
of masses of the low-lying hadronic states.  In the papers of
70-80-ies it was adopted that the perturbaive interaction constant
is comparatively small (e.g., $\alpha_s(1 GeV) \approx 0.3)$, so
that it is enough to restrict oneself by the first-order terms in
$\alpha_s$ and sometimes even disregard perturbative effects in
the region of masses larger than 1 GeV. At present it is clear
that $\alpha_s$ is considerably larger ($\alpha_s(1 GeV) \sim
0.6$). In a number of cases there appeared the results of
perturbative calculations in order $\alpha^2_s$ and $\alpha^3_s$.
New, more precise experimental data at low energies had been
obtained. Thereby, on one hand, it is necessary, and on the other
it appears to be possible to compare QCD with the experiment in
the low energy region on a higher level of precision.  The results
of such a comparison are presented in this paper.

In Section 3 I define condensates, describe their properties and
give numerical values which were obtained previously. In Section
4, the data on hadronic decays of $\tau$-lepton are compared with
theoretical expectations obtained on the basis of the operator
product expansion in QCD with the account of perturbative terms up
to $\alpha^3_s$. The values of condensates and the coupling
constant $\alpha_s(m^2_{\tau})$ are obtained. In Section 5,
polarization operator of the vector current of charmed quarks is
analysed in three-loop approximation (i.e., with the account of
the terms $\sim \alpha^2_s$), the value of the charmed quark and
the value of gluonic condensate are found. Section 6 presents our
conclusions.

\vspace{3mm}

\begin{center}
{\bf \large 3. Definition of condensates, their main properties}
\end{center}


In QCD (or in a more general case, in quantum field theory) by
condensates there are called the vacuum mean values $\langle 0
\vert O_i \vert 0 \rangle$  of the local (i.e.  taken at a single
point of space-time) of the operators $O_i(x)$, which arise due to
nonperturbative effects. The latter point is very important and
needs clarification. When determining vacuum condensates one
implies the averaging only over nonperturbative fluctuations.  If
for some operator $O_i$ the non-zero vacuum mean value appears
also in the perturbation theory, it should not be taken into
account in determination of the condensate -- in other words, when
determining condensates the perturbative vacuum mean values should
be subtracted in calculation of the vacuum averages. One more
specification is necessary. The perturbation theory series in QCD
are asymptotic series.  So, vacuum mean operator values may appear
due to one or another summing of asymptotic series.  The vacuum
mean values of such kind are commonly to be referred to vacuum
condensates.

Separation of perturbative and nonperturbative contributon into vacuum mean
values has some arbitrariness. Usually [6,7], this arbitrariness is avoided
by introduction of some normalization point $\mu^2$ ~ ($\mu^2 \sim 1
GeV^2$). Integration over  momenta of virtual quarks and gluons in the
region below $\mu^2$ is referred to condensates, above $\mu^2$ --
to perturbative theory. In such a formulation condensates depend on the
normalization point $\mu$: ~ $\langle 0 \vert O_i \vert 0 \rangle = \langle
0 \vert O_i \vert 0 \rangle _{\mu}$. Other methods for determination of
condensates are also possible (see below).

In perturbation theory, there appear corrections to the condensates as a
series in the coupling constant $\alpha_s(\mu)$:
\be
\langle 0 \vert O_i \vert 0 \rangle_Q = \langle 0 \vert O_i \vert
0 \rangle_{\mu} \sum\limits^{\infty}_{n=0}~ C^{(i)}_n (Q, \mu)
\alpha^n_s (\mu) \ee The running coupling constant $\alpha_s$ at
the right-hand part of (1) is normalized at the point $\mu$. The
left-hand part of (1) represents the value of the condensate
normalized at the point $Q$. Coefficients $C^{(i)}_n(Q, \mu)$ may
have logarithms $ln Q^2/\mu^2$ in powers up to $n$ for
$C^{(i)}_n$. Summing up of the terms with highest powers of
logarithms leads to appearance of the so-called anomalous
dimension of operators, so that in general form it can be written
\be
\langle 0 \vert O_i \vert 0 \rangle = \langle 0 \vert O_i \vert 0
\rangle_{\mu} \Biggl
(\frac{\alpha_s(\mu)}{\alpha_s(Q)} \Biggr
)^{\gamma} \sum\limits^{\infty}_{n=0}~ c^{(i)}_n (Q, \mu) \alpha^n_s (\mu),
\ee
where $\gamma$ - is anomalous dimension (number), and $c^{(i)}_n$ has
already no leading logarithms. If there exist several operators of the given
(canonical) dimension, then their mixing is possible in perturbation theory.
Then the relations (1),(2) become matrix ones.

In their physical properties condensates in QCD have much in
common with condensates appearing in condensed matter physics:
such as superfluid liquid (Bose-condensate) in liquid $^4He$,
Cooper pair condensate in superconductor, spontaneous magnetizaion
in magnetic etc. That is why, analogously to effects in the
physics of condensed  matter, it can be expected that if one
considers QCD at finite temperature $T$, with $T$ increasing at
some $T = T_c$ there will be phase transition and condensates (or
a part of them) will be destroyed. Particularly, such a phenomenon
must hold for condensates responsible for spontaneous symmetry
breaking -- at $T = T_c$ they should vanish and symmetry must be
restored. (In principle, surely, QCD may have a few phase
transition).

Condensates in QCD are divided into two types: conserving and violating
chirality. As is known, the masses of light quarks $u, d, s$ in the QCD
Lagrangian are small comparing with the characteristic scale of hadronic
masses $ M \sim 1 GeV$. In neglecting  light quark masses the QCD Lagrangian
becomes chiral-invariant: left-hand and right-hand (in chirality) light
quarks do not interact with each other, both vector and axial currents are
conserved (except for flavour-singlet axial current, non-conservation of
which is due to anomaly). The accuracy of light quark masses neglect
corresponds to the accuracy of isotopical symmetry, i.e. a few per cent in
the case of $u$ and $d$ quarks and of the accuracy of SU(3) symmetry, i.e.
10-15 \% in the case of $s$-quarks. In the case of condensates violating
chiral symmetry, perturbative vacuum mean values are proportional to light
quark masses and are zero within $m_u = m_d = m_s = 0$. So, such condensates
are determined in the theory much better than those conserving chirality
and, in principle, may be found experimentally with higher accuracy.

Among chiral symmetry violating condensates of the most importance is the
quark condensate $\langle 0 \vert \bar{q} q \vert 0 \rangle$~ ($q = u, d$
are the fields of $u$ and $d$ quarks). $\langle 0 \vert \bar{q} q \vert 0
\rangle$ may be written in the form
\be
\langle 0 \vert \bar{q} q \vert 0 \rangle = \langle 0 \vert \bar{q}_L q _R
+ \bar{q}_R q_L \vert 0 \rangle
\ee
where $q_L, q_R$ are the fields of left-hand and right-hand (in chirality)
quarks. As follows from (3), the non-zero value of quark condensate means
the transition of left-hand quark fields into right-hand ones and its not a
small value would mean to chiral symmetry violation in QCD. (If chiral
symmetry is not violated, then at small $m_u, m_d$ ~~ $\langle 0 \vert
\bar{q} q \vert 0 \rangle \sim m_u, m_d$). By virtue of isotopical
invariance
\be
\langle 0 \vert \bar{u} u \vert 0 \rangle = \langle 0 \vert
\bar{d} d \vert 0 \rangle \ee For quark condensate there holds the
Gell-Mann-Oakes-Renner relation [8]
\be
\langle 0 \vert \bar{q} q \vert 0 \rangle = - \frac{1}{2}~ \frac{m^2_{\pi}
f^2 _{\pi}}{m_u + m_d}
\ee
Here $m_{\pi}, f_{\pi}$ are the mass and constant of $\pi^+$-meson decay
($m_{\pi} = 140 MeV, ~ f_{\pi} = 131 MeV$), ~ $m_u$ and $m_d$ are the masses
of $u$ and $d$-quarks. Relation (5) is obtained in the first order in $m_u,
m_d, m_s$ (for its derivation see, e.g. [9]). To estimate the value of quark
condensate one may use the values of quark masses $m_u = 4.2 MeV$, ~ $m_d =
7.5 MeV$ [9]. (These values were suggested by Weinberg [10], within the
errors they coincide with other estimates -- see, for example, [11]).
Substituting these values into (5) we get
\be
\langle 0 \vert \bar{q} q \vert 0 \rangle = - (243 MeV)^3
\ee
The value (6) has characteristic hadronic scale. This shows that chiral
symmetry which is fulfilled with a good accuracy in the light quark
lagrangian ($m_u, m_d/M \sim 0.01$, $M$ - is hadronic mass scale, $M \sim
0.5 - 1 GeV$), is spontaneously violated on hadronic state spectrum.

An other argument in the favour of spontaneous violation of chiral symmetry
in QCD is the existence of massive baryons. Indeed, in the chiral-symmetrical
theory all fermionic states should be either massless or parity-degenerated.
Obviously, baryons, in particular, nucleon do not possess this property. It
can be shown [12, 9], that both these phenomena -- the presence of the
chiral symmetry violating quark condensate and the existence of massive
baryons are closely connected with each other. According to the Goldstone
theorem, the spontaneous symmetry violation leads to appearance of massless
particles in the physical state spectrum -- of Goldstone bosons. In QCD
Goldstone bosons can be identified with a $\pi$-meson triplet within $m_u,
m_d \to 0$, ~ $m_s \not= 0$ (SU(2)-symmetry) or with an octet of
pseudoscalar mesons ($\pi$, $K, \eta$) within the limit $m_u, m_d, m_s \to
0$ (SU(3)-symmetry). The presence of Goldstone bosons in QCD makes it
possible to formulate the low-energy chiral effective theory of strong
interactions (see reviews [13], [14], [9]).

Quark condensate may be considered as an order parameter in QCD
corresponding to spontaneous violation of the chiral symmetry. At the
temperature of restoration of the chiral symmetry $T = T_c$
it must vanish. The investigation of the temperature dependence of quark
condensate in chiral effective theory [15] (see also the review [9]) shows
that  $\langle 0 \vert \bar{q} q \vert 0 \rangle$ vanishes
at $T = T_c \approx 150-200 MeV$. Similar indications were obtained also in
the lattice calculations [16].

Thus, the quark condensate: 1)has the lowest dimensions (d=3) as compared
with other condensates in QCD; 2) determines masses of usual (nonstrange)
baryons; 3) is the order parameter in the phase transition
between the phases of violated and restored chiral symmetry. These
three facts determine its important role in the low-energy hadronic physics.

Let us estimate the accuracy of numerical value of (6). The
Gell-Mann-Oakes-Renner relation is derived up to correction terms
linear in quark masses. In the chiral effective theory one
succeeds in estimating the correction terms and,thereby, the
accuracy of equation (5) appears of order 10\%. However, it is not
a single origin of errors in determination of quark condensate
value. The quark condensate, as well as quark masses depend on the
normalization point and have anomalous dimensions equalling to
$\gamma_m = - \gamma_{\bar{q}q} = \frac{4}{9}$. In the mass values
taken above the normalization point $\mu$ was not fixed exactly
(in fact, it was taken $\mu \sim 1 GeV$). In addition, the
accuracy of the above taken value $m_u + m_d = 11.7 MeV$ which
enters (5) seems to be of order $10-20\%$. The value of the quark
condensate may be also found from the sum rules for proton mass.
The analysis made [17] gave for it a value very close to (6) (with
the 3$\%$ difference) at the normalization point $\mu = 1 GeV$.
The accuracy of these sum rules seems to be of order 10-15$\%$.
Concludingly, it may be believed, that the value of the quark
condensate is given by (6) at the normalization point $\mu = 1
GeV$ with the 10-20$\%$ accuracy.  The quark condensate of strange
quarks is somewhat different from $\langle 0 \vert \bar{u}u \vert
0 \rangle$. In [12] it was obtained
\be
\langle 0 \vert \bar{s} s \vert 0 \rangle/ \langle 0 \vert \bar{u} u \vert 0
\rangle = 0.8 \pm 0.1
\ee
The next in dimension (d = 5) condensate which violates chiral symmetry is
quark gluonic one:
\be
-g \langle 0 \vert \bar{q} \sigma_{\mu \nu} \frac{\lambda^n}{2} G^n_{\mu
\nu} q \vert 0 \rangle \equiv m^2_0 \langle 0 \vert \bar{q} q \vert \rangle
\ee
Here $G^n_{\mu \nu}$ - is the gluonic field strength tensor, $\lambda^n$
- are the Gell-Mann matrices, $\sigma_{\mu \nu} = (i/2)(\gamma_{\mu}
\gamma_{\nu} - \gamma_{\nu}\gamma_{\mu}$. The value of the parameter $m^2_0$
was found in [18] from the sum rules for baryonic resonances
\be
m^2_0 = 0.8 GeV^2
\ee
Consider now condensates conserving chirality. Of fundamental role here is
the gluonic condensate of the lowest dimension:
\be
\langle 0 \vert \frac{\alpha_s}{\pi} G^n_{\mu \nu} G^n_{\mu \nu}
\vert 0\rangle \ee Due to that the gluonic condensate is
proportional to the vacuum mean value of the trace of the
energy-momentum tensor $\theta_{\mu \nu}$ its anomalous dimension
is zero. The existence of gluonic condensate had been first
indicated by Shifman, Vainshtein, and Zakharov [5]. They had also
obtained from the sum rules for charmonium its numerical value:
\be
\langle 0 \vert \frac{\alpha_s}{\pi} G^n_{\mu \nu} G^n_{\mu \nu}
\vert 0\rangle =0. 012 GeV^4 \ee As was shown by the same authors,
the nonzero and positive value of gluonic condensate mean, that
the vacuum energy is negative in QCD: vacuum energy density in QCD
is given by $\varepsilon = -(9/32) \langle 0 \vert (\alpha_s/\pi)
G^2 \vert 0 \rangle$. Therefore, if quark is embedded into vacuum,
this results in its excitation, i.e, in increasing of energy. In
this way, the explanation of the bag model could be obtained in
QCD: in the domain around quark there appears an excess of energy,
which is treated as the energy density $B$ in the bag model.
(Although, the magnitude of $B$, does not,probably, agree with the
value of $\varepsilon$ which follows from (11)). In ref.[5]
perturbative effects were taken into account only in the order
$\alpha_s$, the value for $\alpha_s$ being taken twice as smaller
as the modern one. Later many attempts were made to determine the
value of gluonic condensate by studying various processes and by
applying various methods.  But the results of different approaches
were inconsistent with each other and with (11) and sometimes the
difference was even very large -- the values of condensate
appeared to be by a few times larger. All of this needs
reanalysation  of $\langle 0 \vert \frac{\alpha_s}{\pi} G^2 \vert
0 \rangle$  determination basing on contemporary values which will
be done in Sections 4,5.

The d=6 gluonic
condensate is of the form
\be
g^3 f^{abc} \langle 0 \vert G^a_{\mu \nu} G^b_{\nu \lambda}
G^c_{\lambda \mu} \vert 0 \rangle, \ee $(f^{abc}$ - are structure
constants of SU(3) group). There are no reliable methods to
determine it from experimental data. There is only an estimate
which follows from the method of deluted instanton gas [19]:
\be
g^3 f^{abc} \langle 0 \vert G^a_{\mu \nu} G^b{\nu \lambda} G^c_{\lambda \mu}
\vert 0 \rangle = \frac{4}{5} (12 \pi^2) \frac{1}{\rho^2_c} \langle 0 \vert
\frac{\alpha_s}{\pi} G^2_{\mu \nu} \vert 0 \rangle,
\ee
where $\rho_c$ is the instanton effective radius in the given model (for
estimaion one may take $\rho_c \sim (1/3 - 1/2) fm)$.

The general form of d=6 condensates is as follows:
\be
\alpha_s \langle 0 \vert \bar{q}_i O_{\alpha} q_i \cdot \bar{q}_k
O_{\alpha} q_k \vert 0 \rangle \ee where $q_i, q_k$ are quark
fields of $u, d, s$ quarks, $O_{\alpha}$ - are Dirac and $SU(3)$
matrices. Following [5], Eq.(14) is usually factorized: in the sum
over intermediate state  in all channels (i.e,  if necessary,
after Fierz-transformation) only vacuum state is taken into
account. The accuracy of such approximation $\sim 1/N^2_c$, where
$N_c$ is the number of colours i.e.$\sim 10\%$. After factorizaion
Eq.(14) reduces to
\be
\alpha_s \langle 0 \vert \bar{q} q \vert 0 \rangle^2
\ee
The anomalous dimension of (15) is 1/9 and it can be approximately put
to be zero. And finally, d=8 quark condensates assuming factorization reduce
to
\be
\alpha_s \langle 0 \vert \bar{q} q \vert 0 \rangle  \cdot m^2_0
\langle 0 \vert  \bar{q} q \vert 0 \rangle \ee (The notaion of (8)
is used). It should be noted, however, that the factorization
procedure in the d=8 condensate case is uncertain. For this
reason, it is necessary to require their contribution to be small.

Let us
also dwell on one more type of condensates - those, induced by external
fields.  The meaning of such condensates can be easily understood by
comparing with analogous phenomena in the physics of condenced media. If the
above considered condensates can be compared, for instance with
ferromagnetics, where magnetization is present even in the absence of
external magnetic field, condensates induced by external field are similar
to dia- or paramagnetics. Consider the case of the constant external
electromagnetic field $F_{\mu \nu}$. In its presence there appears a
condensate induced  by external field (in the linear approximation in
$F_{\mu \nu}$):
\be
\langle 0 \vert \bar{q} \sigma_{\mu \nu} q \vert 0 \rangle_F = e_q
\chi F_{\mu \nu} \langle 0 \vert \bar{q} q \vert 0 \rangle \ee As
was shown in ref.[20], in a good approximation  $\langle 0 \vert
\bar{q} \sigma_{\mu \nu} q \vert 0 \rangle_F$ is proportional to
$e_q$ - the charge of quark $q$. Induced, by the field vacuum
expectation value $\langle 0 \vert \bar{q} \sigma_{\mu \nu} q
\vert 0 \rangle_F$ violates chiral symmetry.  So, it is natural to
separate $\langle 0 \vert \bar{q}q \vert 0 \rangle$ as a factor in
Eq.(17).  The universal quark flavour independent quantity $\chi$
is called magnetic susceptibility of quark condensate. Its
numerical value had been found in [21] using a special sum rule:
\be
\chi = -(5.7 \pm 0.6) GeV^2
\ee
Another example is external constant axial isovector field $A_{\mu}$ the
interaction of which with light quarks is described by Lagrangian
\be
L^{\prime} = (\bar{u} \gamma_{\mu} \gamma_5 u - \bar{d} \gamma_{\mu}
\gamma_5 d) A_{\mu}
\ee
In the presence of this field there appear induced by it condensates:
\be
\langle 0 \vert \bar{u} \gamma_{\mu} \gamma_5 u \vert 0 \rangle_A
= - \langle 0 \vert \bar{d} \gamma_{\mu} \gamma_5 \vert 0 \rangle
_A = f^2_{\pi} A_{\mu} \ee where $f_{\pi} = 131 MeV$ is the
constant of $\pi \to \mu \nu$ decay. The right-hand part of
eq.(20) is obtained assuming $m_u, m_d \to 0,$~ $m^2_{\pi} \to 0$
and follows directly from consideration of the polarization
operator of axial currents $\Pi^A_{\mu \nu}(q)$  in the limit $q
\to 0$, when nonzero contribution into $\Pi^A_{\mu \nu}(q)_{q \to
0}$ emerges only from one-pion intermediate state. The equality
(20) was used to calculate the axial coupling constant in
$\beta$-decay $q_A$ [22]. An analogous to (20) relation holds in
the case of octet axial field. Of special interest is the
condensate induced by singlet (by flavours) constant axial field
\be
\langle 0 \vert \gamma^{(0)}_{\mu 5} \vert 0 \rangle = 3 f^2_0
A^{(0)}_{\mu} \ee
\be
j^{(0)}_{\mu 5} = \bar{u} \gamma_{\mu} \gamma_5 u + \bar{d}
\gamma_{\mu} \gamma_5 d +  \bar{s} \gamma_{\mu} \gamma_5 s \ee and
Lagrangian of interaction with external field has the form
\be
L^{\prime} = j^{(0)}_{\mu 5} A^{(0)}_{\mu} \ee Constant $f_0$
cannot be calculated by the method used when deriving eq.(20),
since singlet axial current is not conserved by virtue of anomaly
and the singlet pseudoscalar meson  $\eta^{\prime}$ is not
Goldstone one. Constant $f^2_0$ is proportional to topologicaal
susceptibility of vacuum [23]
\be
f^2_0 = \frac{4}{3} N^2_f \chi^{\prime} (0),
\ee
where $N_f$ is the number of light quarks, $N_f = 3$, and the topological
susceptibility of the vacuum $\chi(q^2)$ is defined as
\be
\chi(q^2) = i \int~ d^4 xe^{iqx} \langle 0 \vert T {Q_5(x),~ Q_5(0)} \vert 0
\rangle
\ee
\be
Q_5(x) = \frac{\alpha_s}{8 \pi} G^n_{\mu \nu} (x) G^n_{\mu \nu} (x)
\ee
Using the QCD sum rule, one may relate $f^2_0$ with the part of proton spin
$\Sigma$, carried by quarks in polarized $ep$ (or $\mu p$) scattering [23].
The value of $f^2_0$ was found from the selfconsistency condition of obtained
sum rule (or from the experimental value of $\Sigma$):
\be
f^2_0 = (2.8 \pm 0.7) \cdot 10^{-2} GeV^2
\ee
The related to it value of the derivative at $q^2 = 0$ of vacuum topological
susceptibility $\chi^{\prime}(0)$, (more precisely, its
nonperturbative part) is equal to:
\be
\chi^{\prime} (0) = (2.3 \pm 0.6) \cdot 10^{-3} GeV^2
\ee
The value $\chi^{\prime} (0)$ is of essential interest for studying
properties of vacuum in QCD.

\vspace{8mm}

\centerline{\bf \large 4. Test of QCD at low energies on the basis
of $\tau$-decay data.}

\vspace{1mm}

{\bf \large Determination of $\alpha_s(m^2_{\tau})$
and of condensate values.}

\vspace{5mm}

Recently, collaborations ALEPH [24], OPAL [25] and CLEO [26] had
measured with a good accuracy the relative probability of hadronic
decays of $\tau$-lepton $R_{\tau} = B(\tau \to \nu_{\tau} +
hadrons)/B(\tau \to \nu_{\tau} e\overline{\nu}_e)$, the vector $V$
and axial $A$ spectral functions. Below I present the results of
the theoretical analysis of these data basing on the operator
product expansion (OPE) in QCD [27, 28]. In the perturbation
theory series the terms up to $\alpha^3_s$ will be taken into
account, in OPE -- the operators up to dimension 8.

Consider the polarization operator of hadronic currents
\be
\Pi^J_{\mu \nu} = i~ \int~ e^{iqx} \langle T J_{\mu} (x) J_{\nu}
(0)^{\dag} \rangle dx = (q_{\mu} q_{\nu} - q_{\mu \nu} q^2)
\Pi^{(1)}_J (q^2) + q_{\mu} q_{\nu} \Pi^{(0)}_J (q^2), \ee $$
\mbox{where} ~~~~ J = V,A; ~~~ V_{\mu} = \bar{u} \gamma_{\mu} d,
~~~ A_{\mu} = \bar{u} \gamma_{\mu} \gamma_5 d. $$ The spectral
functions measured in $\tau$-decay are imaginary parts of
$\Pi^{(1)}_J(s)$ and $\Pi^{(0)}_J(s)$, ~ $s = q^2$

\be
v_1/a_1(s) = 2\pi Im \Pi^{(1)}_{V/A} (s + i 0), ~~~ a_0(s) = 2 \pi Im
\Pi^{(0)}_A (s + i0)
\ee
Functions $\Pi^{(1)}_V(q^2)$ and $\Pi^{(0)}_A(q^2)$ are analytical functions
in the $q^2$ complex plane  with a cut along the right-hand semiaxis starting
from $4 m^2_{\pi}$ for $\Pi^{(1)}_V(q^2)$ and $9m^2_{\pi}$ for $\Pi^{(0)}_A(
q^2)$. Function $\Pi^{(1)}_A(q^2)$ has kinematical pole at $q^2 = 0$. This
is a specific feature of QCD following from chiral symmetry within massless
$u$ and $d$ quarks and from its spontaneous violation. The kinematical pole
appears due to one-pion state contribution into $\Pi_A(q)$, which has
the form [27]
\be
\Pi^A_{\mu \nu}(q)_{\pi} = -\frac{f^2_{\pi}}{q^2} (q_{\mu} q_{\nu}
- q_{\mu \nu} q^2) - \frac{m^2_{\pi}}{q^2} q_{\mu} q_{\nu}
\frac{f^2_{\pi}}{q^2 - m^2_{\pi}} \ee Consider first the ratio of
the total probability of hadronic decays of $\tau$-lepons into
states with zero strangeness to the probability of $\tau \to
\nu_{\tau} e \overline{\nu}_e$. This ratio is given by the
equality [29]
$$
R_{\tau, V+A} = \frac{B(\tau \to \nu_{\tau} + hadrons_{S=0})}{B(\tau \to
\nu_{\tau} e\bar{\nu}_e)} =
$$
\be
= 6 \vert V_{ud} \vert^2 S_{EW}~ \int\limits^{m^2_{\tau}}_{0}~
\frac{ds}{m^2_{\tau}} \Biggl ( 1 - \frac{s}{m^2_{\tau}} \Biggr )^2 \Biggl [
\Biggl ( 1 + 2 \frac{s}{m^2_{\tau}} \Biggr ) (v_1 + a_1 +a_0)(s) - 2
\frac{s}{m^2_{\tau}} a_0(s) \Biggr ]
\ee
where $\vert V_{ud} \vert = 0.9735 \pm 0.0008$ is the matrix element of
the Kabayashi-Maskawa matrix, $S_{EW} = 1.0194 \pm 0.0040$ is the
electroweak correction [30]. Only one-pion state is practically contributing
to the last term in (32) and it appears to be small:
\be
\Delta R^{(0)}_{\tau} = - 24 \pi^2 \frac{f^2_{\pi} m^2_{\pi}}{m^4_{\tau}} =
- 0.008
\ee
Denote
\be
\omega(s) \equiv v_1 + a_1 +a_6 = Im [\Pi^{(1)}_V(s) + \Pi{(1)}_A(s) +
\Pi^{(0)}_A(s) ] \equiv 2 \pi Im \Pi(s)
\ee
As follows from eq.(31),
$\Pi(s)$ has no kinematical pole, but only right-hand cut. It is convenient
to transform the integral in eq.(32) into that over the circle of radius
$m^2_{\tau}$ in the complex $s$ plane [31]-[33]:
\be
R_{\tau,\, V+A} = 6\pi i |V_{ud}|^2 S_{EW}
\oint_{|s|=m_\tau^2}\!{ds\over m_\tau^2} \left( 1-{s\over
m_\tau^2} \right)^2 \left( 1+2 {s\over m_\tau^2}\right) \Pi (s) +
\Delta R_\tau^{(0)} \label{35} \ee Calculate first the
perturbative contribution into eq.(35). To this end, use the Adler
function $D(Q^2)$:
\be
D(Q^2) \, \equiv \, - 2\pi^2 \,{ d\Pi(Q^2)\over d\ln{Q^2}}
\,=\,\sum_{n\ge 0} K_n a^n
 \; , \qquad a\equiv {\alpha_s \over \pi}\; , \qquad  Q^2\equiv
 -s,\label{36} \ee
the perturbative expansion of which is known up to terms $\sim
\alpha^3_s$. In $\overline{MS}$ regularization scheme $K_0 = K_1 =
1$,~~ $K_2 = 1.64$ [34], $K_3 = 6.37$ [35] for 3 flavours and for
$K_4$ there is the estimate $K_4 = 25 \pm 25$ [36]. The
renormgroup equation yields
\be
{d a \over d \ln{Q^2}} \, =\, -\beta(a) \,=\, - \sum_{n\ge 0}
\beta_n a^{n+2}  \qquad  \Rightarrow
 \qquad \ln{Q^2\over \mu^2}\, = \,-\, \int_{a(\mu^2)}^{a(Q^2)}
 {da\over \beta(a)},
\label{37}
\ee
in the $\overline{MS}$ scheme for three flavours
$\beta_0 = 9/4$,~$\beta_1 = 4$, ~ $\beta_2 = 10.06$, ~ $\beta_3 =
47.23$ ~ [37, 38]. Integrating over eq.(36) and using eq.(38) we
get
\be
\Pi(Q^2)\,=\,{1\over 2\pi^2} \int_{a(\mu^2)}^{a(Q^2)} D(a)
{da\over \beta(a)} \label{38}
\ee

Put $\mu^2 = m^2_{\tau}$ and choose some (arbitrary) value
$a(m^2_{\tau})$. With the help of eq.(37)  one may determine then
$a(Q^2)$ for any $Q^2$ and by analytical continuation for any $s$
in the complex plane. Then, calculating (38) find $\Pi(s)$ in the
whole complex plane. Substitution of $\Pi(s)$ into eq.(35)
determines $R_{\tau}$ for the given $a(m^2_{\tau})$ up to power
corrections. Thereby, knowing $R_\tau$ from experiment it is
possible to find the corresponding to it $a(m^2_\tau)$. Note, that
with such an approach there is no need to expand the nominator in
eqs.(37), (38) in the inverse powers of $ln Q^2/\mu^2$.
Particularly, there is no expansion on the right-hand semiaxis in
powers of the parameter $\pi/ln (Q^2/\mu^2)$, which is not small
in the investigated region of $Q^2$. Advantages  of transformation
of the integral over the real axis (32) in the contour integral
are the following.  It can be expected that the applicability
region of the theory presented as perturbation theory (PT) +
operator expansion (OPE) in the complex $s$-plane is off the
shadowed region in Fig.1. It is evident that at positive and
comparatively small $s$ PT+OPE do not work. At negative $s = -Q^2$
in $\alpha_s$ order a nonphysical pole appears, in higher orders,
according with (9) it is replaced by a nonphysical cut, which
starts from the point $-Q^2_0$, determined by the formula


\begin{figure}[tb]
\hspace{30mm} \epsfig{file=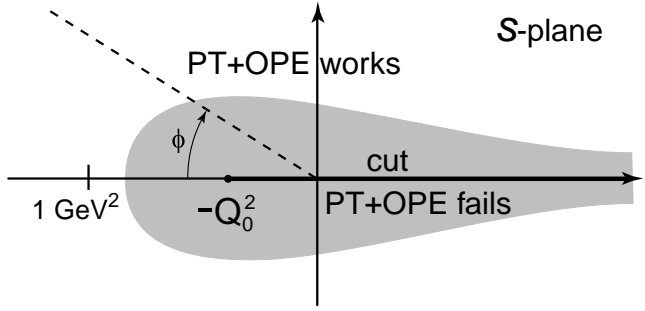, width=85mm} \caption{ The
applicability region of PT and OE in the complex plane $s$. In the
dashed region PT + OE does not work.}
\end{figure}
\be
\ln{Q_0^2\over \mu^2}\, = \,-\, \int_{a(\mu^2)}^\infty {da\over
\beta(a)}\label{39} \ee Integration over the contour allows one to
obviate the dashed region in Fig.1 (except for the vicinity of the
positive semiaxis, the contribution of which, is suppressed  by
the factor $(1 - \frac{s}{m^2_{\tau}})^2$ in eq.(6)), i.e. to work
in the applicability region of PT+OPE. The OPE terms, i.e., power
corrections to polarization operator, are given by the formula
(5):
$$ \Pi(s)_{nonpert}=\sum_{n\ge 2} {\left<O_{2n}\right>\over
(-s)^n} \left( 1+ c_n {\alpha_s\over \pi} \right)
 {\alpha_s\over 6 \pi\, Q^4} \left< G_{\mu\nu}^aG_{\mu\nu}^a\right>\left( 1+
{7\over 6} {\alpha_s\over \pi} \right)+$$
\be
+  {128\over 81\, Q^6}\, \pi\alpha_s \left<\bar{q}q\right>^2
\left[ 1 + \left({29\over 24} + {17\over 18}\ln{Q^2\over \mu^2}
\right){\alpha_s\over \pi}
 \right] + {\left<O_8\right>\over Q^8}  \label{40}
\ee
($\alpha_s$-corrections to the 1-st and 2-d terms in eq.(39)
were calculated in [39] and [40], respectively). Contributions of
the operator with $d=2$ proportional to $m^2_u$, $m^2_d$ and of
the condensate $2(m_u + m_d)\langle 0 \vert \bar{q}q \vert 0
\rangle$ are neglected. (The latter is of an order of magnitude
smaller than the gluonic condensate contribution). When
calculating the d=6  term, factorization hypothesis was used. It
can be readily seen that d=4 condensates (up to small $\alpha_s$
corrections) give no contribution into the integral over contour
eq.(35). The contribution from the condensate $\langle O_8
\rangle$ may be estimated as $\vert \langle O_8 \rangle \vert <
10^{-3} GeV^8$ and appears to be negligibly small. $R_{\tau,V+A}$
may be represented as
$$ R_{\tau, V+A}  = 3|V_{ud}|^2 S_{EW}\left(
\,1\,+\,\delta_{EW}'\,+\,\delta^{(0)}\,+\,\delta^{(6)}_{V+A} \,
\right) +\Delta R^{(0)}~=$$
\be
= ~3.475 \pm 0.022 \label{41}
\ee
where $\delta^{\prime}_{em} =
(5/12 \pi)\alpha_{em}(m^2_{\tau}) = 0.001$ is electromagnetic
correction [41], $\delta^{(6)}_{A+V} = -(5\pm2)\cdot 10^{-3}$ is
the contribution of d=6 condensate (see below) and $\delta^{(0)}$
is the PT correction. The right-hand part presents the
experimental value obtained as a difference between the total
probability of hadronic decays $R_{\tau} = 3.636 \pm 0.021$ [42]
and the probability of decays in states with the strangeness $S =
-1 ~~R_{\tau,s} = 0.161 \pm 0.007$ [43, 44].  For perturbative
correction it follows from eq.(41)
\be \delta^{(0)} = 0.206 \pm 0.010 \label{42}\ee Employing the
above described method in ref.[28] the constant
$\alpha_s(m^2_{\tau})$  was found from (42)
\be
\alpha_s(m^2_{\tau}) = 0.355 \pm 0.025  \label{43}\ee The
calculation was made with the account of terms $\sim
\alpha^3_{\tau}$, the estimate of the effect of the terms
$\sim\alpha^4_s$ is accounted for in the error. May be, the error
is underestimated (by 0.010-0.015), since the theoretical and
experimental errors were added in quadratures.

I determine now the values of condensates basing on the data [24] -[26] on
spectral functions. It is convenient first to consider the difference
$\Pi_V-\Pi_A$, which is not contributed by perturbative terms and there
remains only the OPE contribution:
\be
\label{ope1} \Pi_V^{(1)}(s)-\Pi_A^{(1)}(s)\,=\,\sum_{D\ge 4}
\,{O^{V-A}_D \over (-s)^{D/2} } \left( \,1\,+\,c_D {\alpha_s\over
\pi}\, \right)\,\label{44} \ee The gluonic condensates
contribution falls out in the $V-A$ difference and only the
following condensates with d=4,6,8 remain
\be O^{V-A}_4  =  2 \,(m_u +m_d)\,<\bar{q}q>  \; = \; -\, f_\pi^2
m_\pi^2 \label{45}\ee
$$ O^{V-A}_6  =  2\pi \alpha_s \left<\,
(\bar{u}\gamma_\mu\lambda^a d)(\bar{d}\gamma_\mu \lambda^a u) -
(\bar{u}\gamma_5\gamma_\mu\lambda^a d)(\bar{d}\gamma_5\gamma_\mu
\lambda^a u)\, \right>  = $$ \be = -\,{64\pi\alpha_s\over 9}
<\bar{q}q>^2 \label{46} \ee
\be
 O^{V-A}_8  =   8\pi \alpha_s \, m_0^2
<\bar{q}q>^2 \;,  \label{47} \ee   where $m^2_0$ is determined in
eq.(9). In the right-hand of (46),(47)  the factorization
hypothesis was used. Calculation of the coefficients at $\alpha_s$
in eq.(44) gave $c_4 = 4/3$ [39] and $c_6 = 89/48$ [40]. The value
of $\alpha_s(m^2_{\tau})$  (43) corresponds to $\alpha_s(1 GeV^2)
= 0.60$. Thus, if we take for quark condensate at the normlization
point $\mu^2 = 1 GeV^2$ the value (6), then vacuum condensates
with the account of $\alpha_s$-corrections appear to be equal (at
$\mu^2 = 1 GeV^2$):
\be
O_4=-4.22 \cdot 10^{-4}~GeV^4 \label{48}\ee
\be
O_6=-3.75 \cdot 10^{-3}~GeV^6 \label{49}\ee
\be
O_8=2.5 \cdot 10^{-3}~GeV^8 \label{50}\ee (In what follows,
indeces $V-A$ will be omitted and $O_D$ will mean condensates with
the account of $\alpha_s$ corrections).

Our aim is to compare OPE theoretical predictions with experimental data on
$V-A$ structure functions measured in $\tau$-decay and the values of $O_6$
and $O_8$ found from experiment to compare with
eqs.(49),(50).  Numerical values of $O_6$ and $O_8$ (49),(50) do not
strongly differ. This indicates that OPE asymptotic series (44) at $Q^2 = -s
\sim 1 GeV^2$ converge badly and, may be, even diverge and the role of
higher dimension operators may be essential.  Therefore it is necessary:
either to work at larger $Q^2$, where, however, experimental errors
increase, or to improve the series convergence. The most plausible method is
to use Borel transformation. Write for $\Pi^{(1)}_V - \Pi^{(1)}_A$ the
subtractionless dispersion relation
\be
\Pi^{(1)}_V(s)-\Pi^{(1)}_A(s)\,=\,{1\over 2\pi^2}\int_0^\infty
{v_1(t)-a_1(t)\over t-s} \, dt\,+ \,{f_\pi^2\over s} \label{51}
\ee (The last term in the right-hand part is the kinematic pole
contribution). Put $s = s_0^{i \phi}$  ($\phi = 0$ on the upper
edge of the cut) and make the Borel transformation in $s_0$. As a
result, we get the following sum rules for the real and imaginary
parts of (51):
\be
\int_0^\infty
\exp{\!\left({s\over M^2}\cos{\phi}\right)}\cos{\!\left({s\over
M^2}\sin{\phi}\right)} (v_1-a_1)(s)\,{ds\over 2\pi^2} \, = \,
f_\pi^2+\,\sum_{k=1}^\infty (-)^k {\cos{(k\phi)} \,O_{2k+2}\over k!\,
M^{2k}} \label{52}\ee
\be
\int_0^\infty \exp{\!\left({s\over
M^2}\cos{\phi}\right)}\,\sin{\!\left({s\over
M^2}\sin{\phi}\right)} (v_1-a_1)(s)\,{ds\over 2\pi^2 M^2} \, =
\,\sum_{k=1}^\infty (-)^k {\sin{(k\phi)} \,O_{2k+2}\over k!\,
M^{2k+2}} \label{53}\ee The use of the Borel transformation along
the rays in the complex plane has a number of advantages. The
exponent index is negative at $\pi/2 < \phi < 3 \pi/2$. Choose
$\phi$ in the region $\pi/2 < \phi < \pi$. In this region, on one
hand, the shadowed area in fig.1 in the integrals (52),(53) is
touched to a less degree, and on the other hand, the contribution
of large $s$, particularly, $s > m^2_{\tau}$ , where experimental
data are absent, is exponentially suppressed. At definite values
of $\phi$ the contribution of some condensates vanishes, what may
be also used.  In particular, the condensate $O_8$ does not
contribute to (52) at $\phi = 5 \pi/6$ and to (53) at $\phi = 2
\pi/3$, while  the contribution of $O_6$ vanishes at $\phi = 3
\pi/4$. Finally, a well known advantage of the Borel sum rules is
factorial suppression of higher dimension terms of OPE. Figs.2,3
presents the results of the calculations of left-hand parts of
eqs.(52),(53) on the basis of the ALEPH [24] experimental data
comparing with OPE predictions -- the right-hand part of these
equations.


\begin{figure}[tb]
\hspace{0mm} \epsfig{file=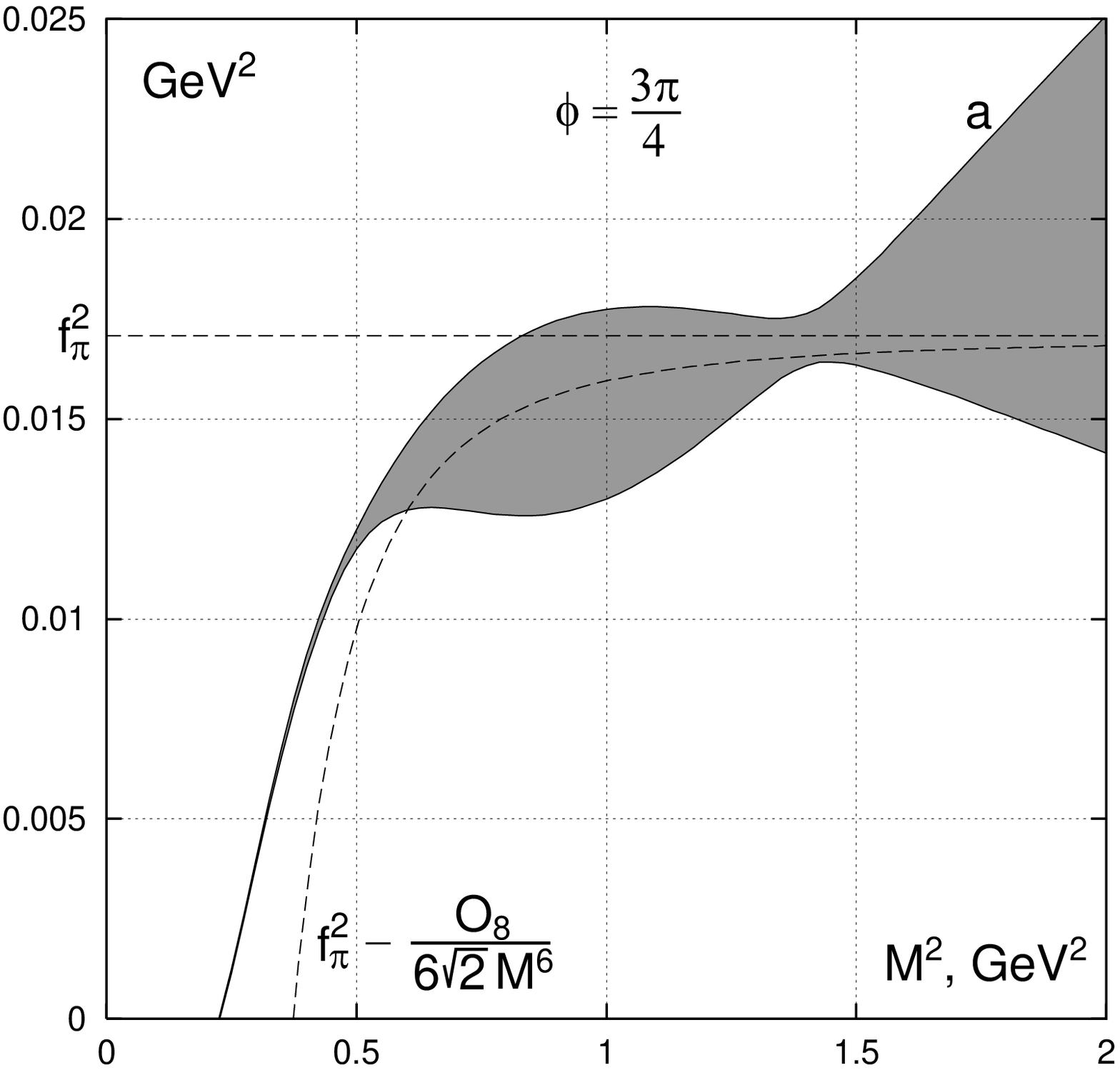, width=70mm} \hspace{5mm}
\epsfig{file=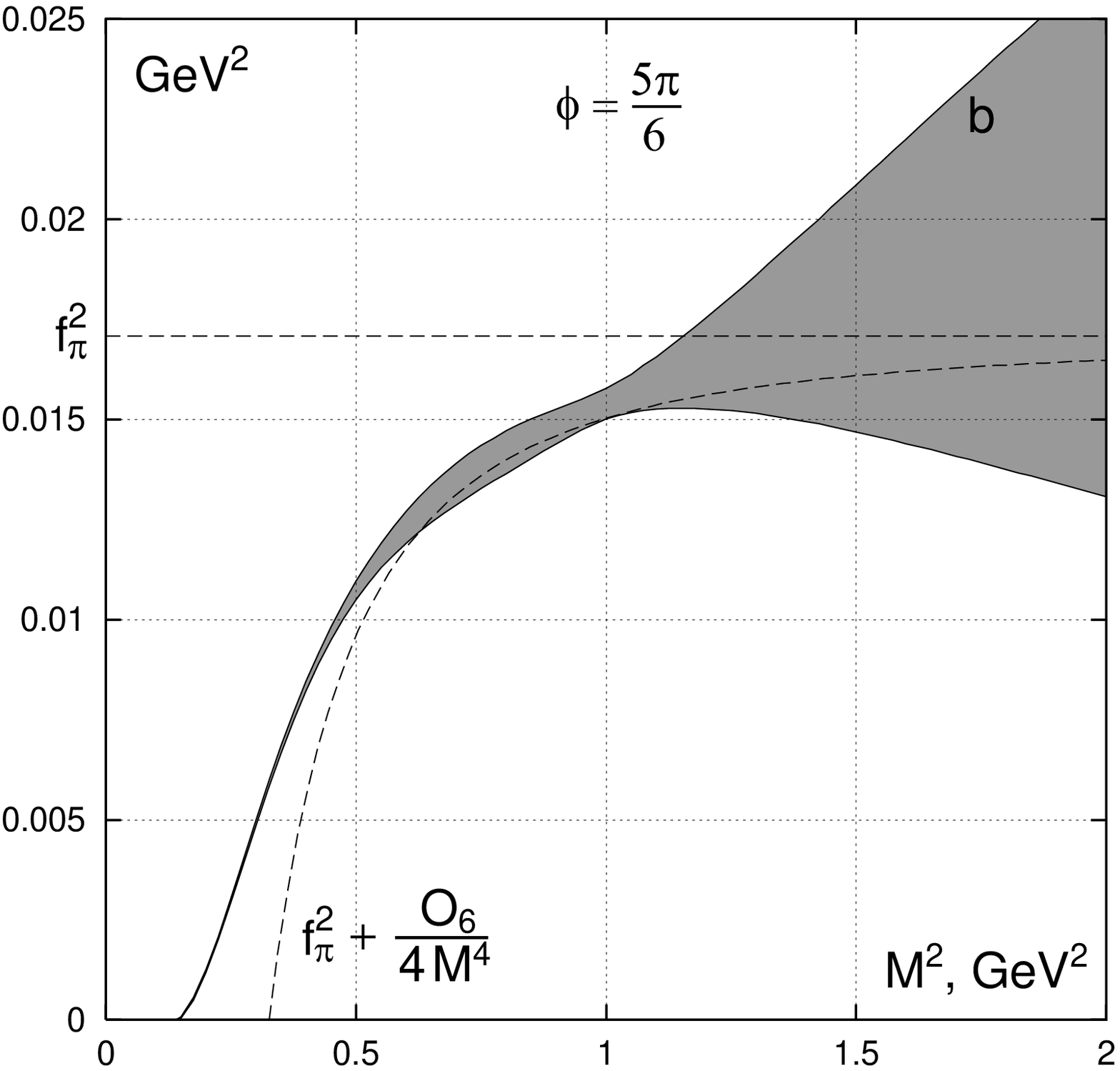, width=70mm} \caption{Eq.(52): the
left-hand part is obtained basing on the experimental data, the
shaded region corresponds to experimental errors; the right-hand
part -- the theoretical one -- is represented by the dotted curve,
numerical values of condensates are taken to be equal to the
central values of eqs.(54),(55); a) $\phi = 3 \pi/4$,~ b) $\phi =
5 \pi/6$.}
\end{figure}
\begin{figure}[tb]
\hspace{0mm} \epsfig{file=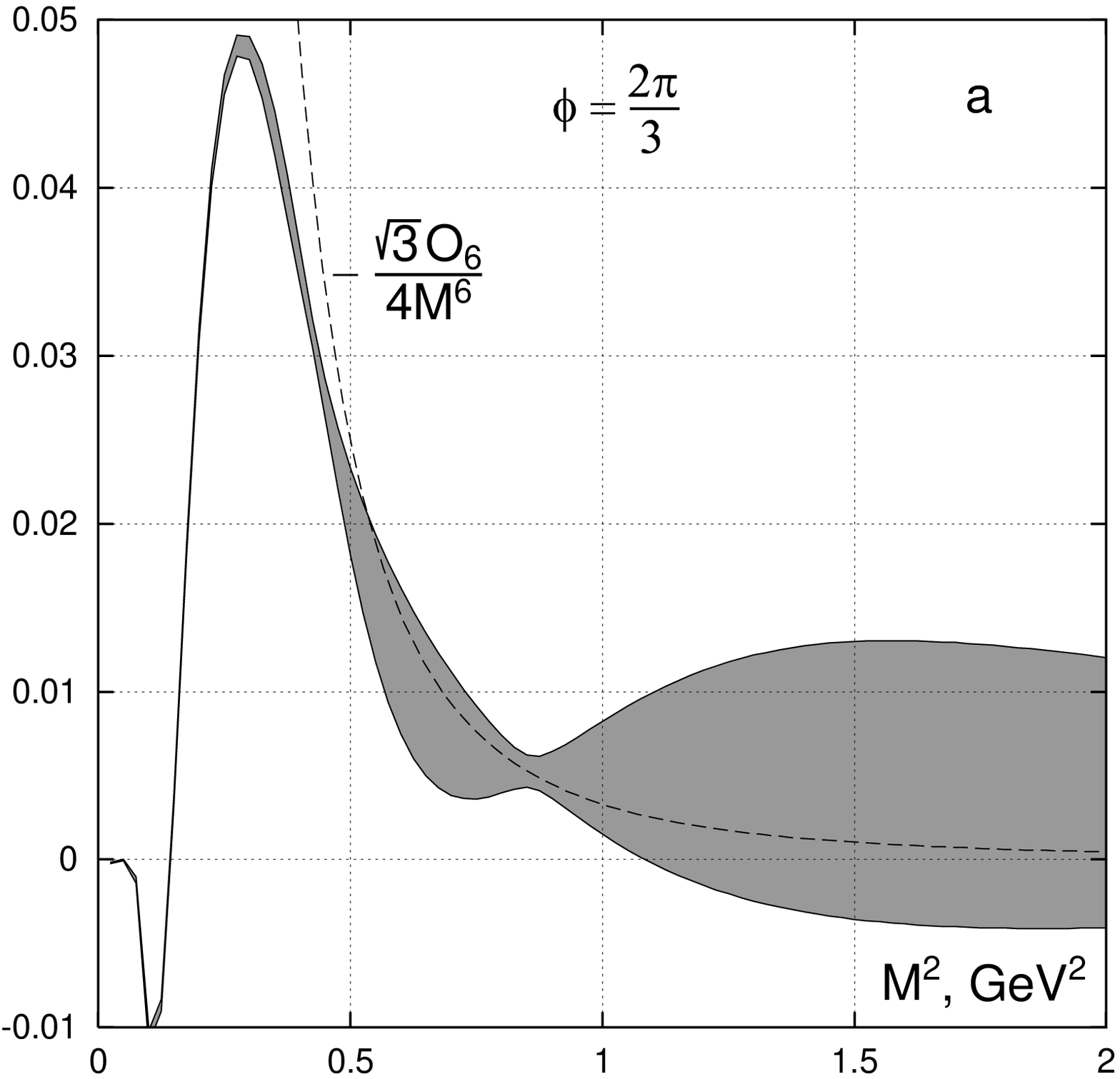, width=70mm} \hspace{5mm}
\epsfig{file=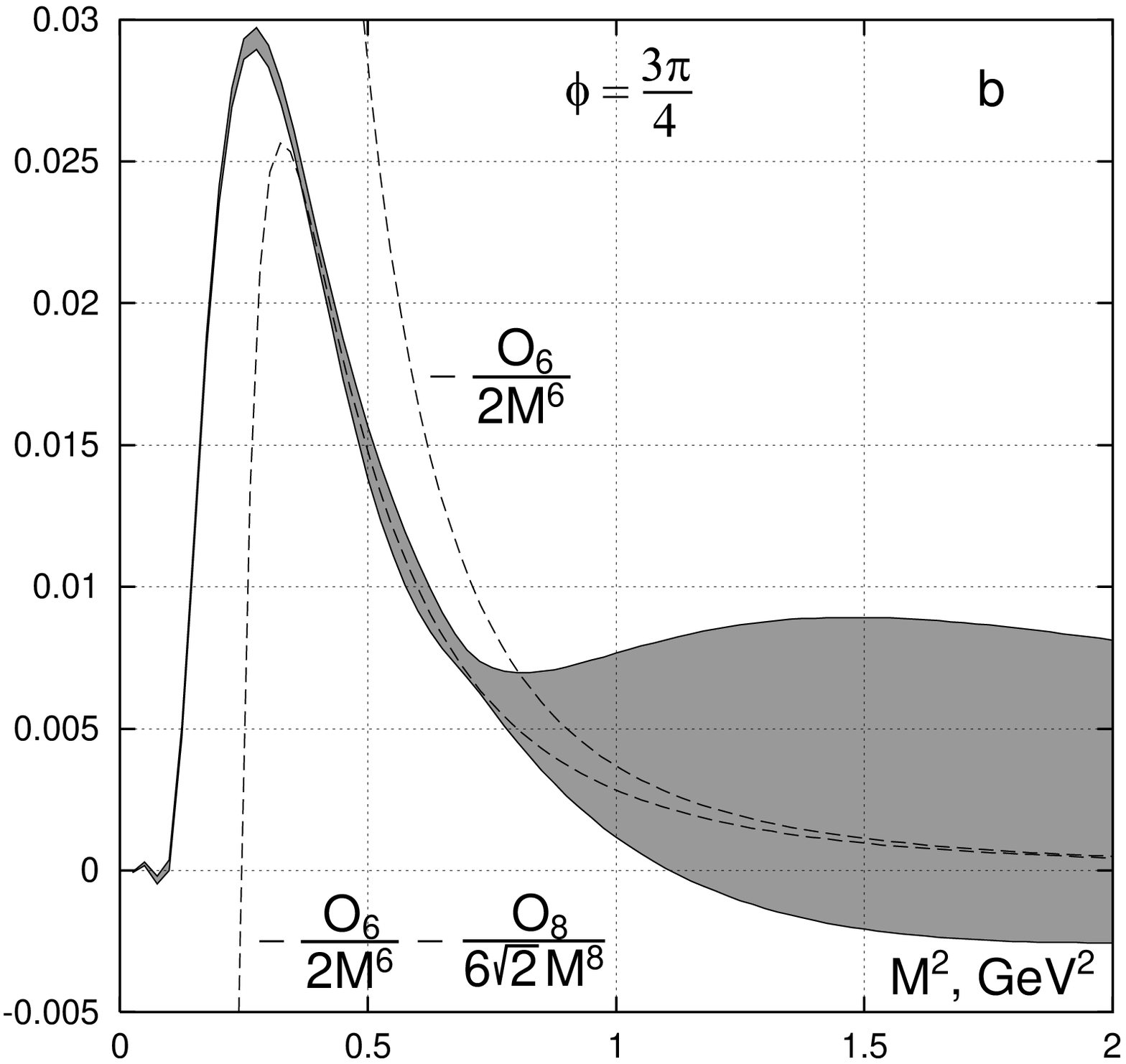, width=70mm} \caption{The same for
eq.(53): a) $\phi = 2 \pi/3$, ~ b) $\phi = 3 \pi/4$. }
\end{figure}
The experimental data are best described at the values [27]
\be
O_6=-(6.8\pm 2.1)\cdot 10^{-3}~GeV^6 \ee
\be
O_8=-(7\pm 4)\cdot 10^{-3}~GeV^8\label{55}\ee When estimating
errors in (54),(55), an uncertainty of higher dimension operator
contribution was taken into account in addition to experimental
errors. (For detils -- see [27]).

As is seen from the figures, at these values of condensates a good
agreemeent with experiment starts rather early -- at $M^2
> 0.5 GeV^2$.  In paper [27] the sum rules for the moments and the
Gaussian sum rules were also considered. All of them agree with
the values of condensates (54), (55), but the accuracy of their
determination is worse. The values (54),(55) are by a factor of
1.5-2 larger than (49),(50). As was discussed above, the accuracy
of (49),(50) is of order 50$\%$. Therefore, the most plausible is
that the real value of condensates $O_6,O_8$ is somewhere close to
the lower edge of errors in (54),(55).

Consider now the polarization operator $\Pi(s)$ defined in (34) and
condensates entering OPE for $\Pi(s)$ (see (40)). In principle, the
perturbative terms contribute to chirality conserving condensates.
If we will follow the separation method of perturbative
and nonperturbative contribution by introducing infrared cut-off [6, 7],
then such a contribution would really appear due to the region of
virtualities smaller than $\mu^2$.  In the present paper, according to [28],
 an another method is exploited, when the $\beta$-function is expanded only
in the number of loops, (see eq.(11) and the text after it) but not in
$1/lnQ^2$.  So, the dependence of condensates on the normalization point
$\mu^2$ is determined only by perturbative corrections, as is seen in (40).
Condensates determined in such a way may be called $n$-loop ones (in the
given case -- 3-loop).  Consider the Borel transformation of the sum
$\Pi(s)_{pert} + \Pi(s)_{nonpert}$ where $\Pi(s)_{pert}$ is given by
eq.(38), and $\Pi(s)_{nonpert}$ -- by eq.(13).  Fig.4 presents the results
of 3-loop calculation for two values of $\alpha_s(m^2_{\tau})$ -- 0.355 and
0.330. The widths of the bands correspond to theoretical error taken to be
equal to the last accounted term $K_3a^2$ in the Adler function (36). (The
same result for the error is obtained if one takes 4 loops in
$\beta$-function and puts $K_4 = 50 \pm 50$). The dotted line corresponds to
the sum  of contributions of gluonic condensate (11) and $O^{V+A}_6$
condensate in (13) with numerical value corresponding to $O^{V-A}_6$ (54).
The dots with errors present experimental data.  (The contribution of the
operators  $d = 4$ and $d = 6$ is given separately in the insert).

It is seen that the curve with $\alpha_s(m^2_{\tau}) = 0.330$ and
condensate contributions can be agreed with experiment, starting
from $M^2 = 1.1 GeV^2$, the agreement being improved at smaller
values $\langle 0 \vert \frac{\alpha_s}{\pi} G^2 \vert 0 \rangle$
than (11). The curve with $\alpha_s(m^2_{\tau}) = 0.355$ with the
account of condensates coincides with experiment only at $M^2 >
1.5 GeV^2$. The same tendency persist for the Borel sum rules
taken along the rays in the $s$ complex plane at various $\phi$.
Fig.5 gives the sum rule for $\phi = 5 \pi/6$. From consideration
of this and of other sum rules there follows the estimation for
gluonic condensate:


\begin{figure}[tb]
\hspace{30mm} \epsfig{file=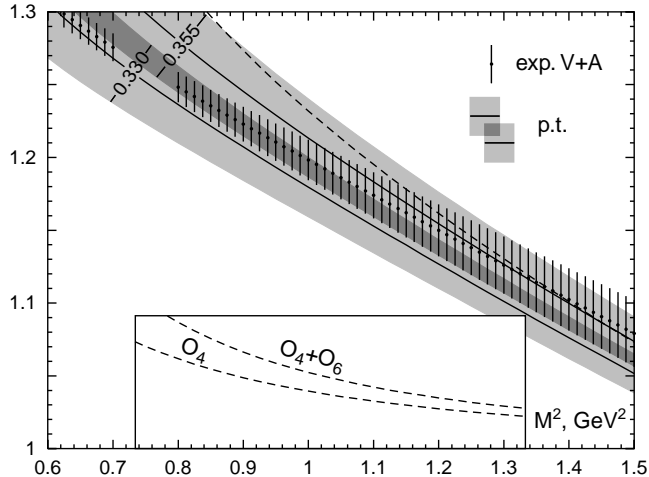, width=85mm} \caption{The
results of the Borel transformation of $V+A$ correlator for two
values $\alpha_s(m^2_{\tau} = 0.355$ and $\alpha_s(m^2_{\tau}) =
0.330$. The widths of the bands correspond to PT errors, dots with
errors -- experimental data. The dotted curve is the sum of the
perturbative contribution at $\alpha_s(m^2_{\tau}) = 0.330$ and
$O_4$,~ $O_6$ condensates.}
\end{figure}


\begin{figure}[tb]
\hspace{30mm} \epsfig{file=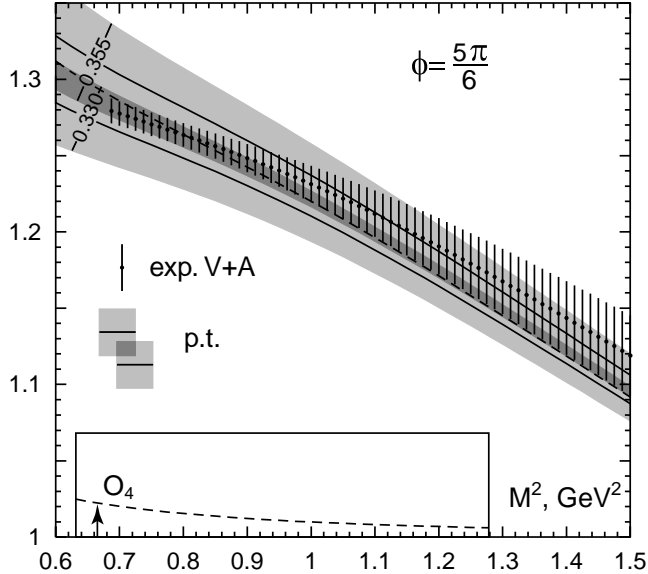, width=85mm} \caption{ The
result of the Borel transformation along the ray at $\phi = 5
\pi/6$. The dotted line corresponds to the central value of
gluonic condensate (56), added to the PT curve with
$\alpha_s(m^2_{\tau}) = 0.330$. }
\end{figure}
\be
\langle 0\mid \frac{\alpha_s}{\pi} G^a_{\mu\nu} G^a_{\mu\nu} \mid
0 \rangle = 0.006 \pm 0.012 ~GeV^4\label{56}\ee The best agreement
of the theory with experiment in the low $Q^2$ region (up to $\sim
2\%$  at $M^2 > 0.8 GeV^2$) is obtained at $\alpha_s(m^2_{\tau}) =
0.330$ which corresponds to $\alpha_s(m^2_z) = 0.118$.

Let us now make  some remarks on modifications of QCD in the low energy
region.

\vspace{3mm}

{\bf 1. Analytical perturbative QCD [45], [46]}. It is assumed
that $\alpha_s(q^2)$ is an analytical function of $q^2$ [45], or,
in a more general case, it is supposed, that the perturbative part
of the polarization operator is an analytical function of $q^2$.
The comparison of this approach with the $\tau$-decay data showed
[28] that in the analytical QCD
\be
\alpha^{anal}_s(m^2_z)=0.140\label{57}\ee what is in a strong
disagreement with the world mean value $\alpha_s(m^2_z) = 0.119
\pm 0.002$.

\vspace{3mm}

{\bf  2. Renormalon summing} leading to the tachion mass
$\lambda^2$ in gluon propagator [47]. The restriction to the
tachion mass
\be
-\lambda^2=0.1\pm 0.15~ GeV^2\label{58}\ee was found from
$\tau$-decay.

\vspace{3mm}

{\bf  3. Instantons}. It was shown [28], that in the dilute
instanton gas appoximation [48] instantons do not practically
affect determination of $\alpha_s(m^2_{\tau})$ and the Borel sum
rules. Their effect, however, appears to be considerable and
strongly dependent on the value of the instanton radius $\rho_c$
in the sum rules obtained by integration over closed contours in
the complex plane $s$ at  the radii of the contours $s < 2 GeV^2$.

\vspace{7mm}

{\bf \large 5. Sum rules for charmonium and gluonic condensate.}

\vspace{5mm}

The value of gluonic condensate had been found by Shifman,
Vainstein, and Zakharov from the sum rules for polarized operator
of vector currents of charmed quarks [5]. But in these
calculations , the constant $\alpha_s$ was taken comparatively
small ($\alpha_s(1 GeV^2 \approx 0.3$;~ $\Lambda^{(3)}_{QCD} = 100
MeV$) and perturbative corrections were taken into account only in
the first order. It is clear now, that $\alpha_s(Q^2)$ in the
region $Q^2  \sim 1 \div 10 GeV^2$ is approximately twice as
large, so that the account of higher order corrections became
necessary. (In what follows I formulate the main results of [49]).

Consider the polarization operator of charmed vector currents
\be
i\int dx \, e^{iqx} \left< \,TJ_\mu (x) J_\nu (0) \, \right>\,=\,
(\,q_\mu q_\nu - g_{\mu\nu}q^2 \,)\, \Pi(q^2) \; , \qquad J_\mu =
\bar{c} \gamma_\mu c \label{59} \ee The dispersion representation
for $\Pi(q^2)$ has the form
\be
R(s)\,=\,4\pi \, {\rm Im} \, \Pi(s+i0) \; , \qquad
\Pi(q^2)\,=\,{q^2\over 4\pi^2}\int_{4m^2}^\infty \,{R(s)\,ds\over
s(s-q^2)} \; , \label{60}\ee where $R(\infty) = 1$ in partonic
model. In approximation of infinitely narrow widths of resonances
$R(s)$ can be written as sums of contributions from resonances and
continuum
\be
\label{rexp} R(s)\,=\,{3 \, \pi \over Q_c^2 \, \alpha_{\rm
em}^2\!(s)}\, \sum_\psi m_\psi \Gamma_{\psi \to
ee}\,\delta(s-m_\psi^2) \,+\,\theta(s-s_0) \label{61}\ee where
$Q_c = 2/3$ is the charge of charmed quarks, $s_0$ - is the
continuum threshold (in what follows $\sqrt{s_0} = 4.6 GeV$), ~~
$\alpha(s)$ - is the running electromagnetic constant,~
$\alpha(m^2_{J/\psi}) = 1/133.6$ Following [5], to suppress the
contribution of higher states and continuum we will study the
polarization operator moments
\be
\label{momdef} M_n(Q^2)  \equiv {4\pi^2\over n!} \left( - {d\over
dQ^2}\right)^n\Pi(-Q^2)= \int_{4m^2}^\infty {R(s)\, ds\over
(s+Q^2)^{n+1}} \label{62}\ee According to (61) the experimental
values of moments are determind by the equality
\be
\label{momexp} M_n(Q^2)\,=\,{27\,\pi\over 4\, \alpha_{\rm
em}^2}\sum_{\psi=1}^6 {m_\psi\Gamma_{\psi\to ee}\over
(m_\psi^2+Q^2)^{n+1}} \,+\,{1\over n (s_0+Q^2)^n} \label{63}\ee It
is reasonable to consider the ratios of moments
$M_{n1}(Q^2)/M_{n2}(Q^2)$ from which the uncertainty due to error
in $\Gamma_{J/\psi \to ee}$ markedly falls out. Theoretical value
for $\Pi(q^2)$ is represented as a sum of perturbative and
nonperturbative contributions. It is convenient to express the
perturbative contribution through $R(s)$, making use of (60),
(62):
\be
R(s)\,=\,\sum_{n\ge 0} R^{(n)}(s,\mu^2)\, a^n(\mu^2) \label{64}\ee
where $a(\mu^2) = \alpha_s(\mu^2)/\pi$. Nowadays, three terms of
expansion in (64) are known: $R^{(0)}$ [51]~ $R^{(1)}$ [52], ~
$R^{(2)}$ [53].  They are represented as functions of quark
velocity $v = \sqrt{1 - 4m^2/s}$, ~ where $m$ - is the pole mass
of quark. Since they are cumbersome, I will not present them here.

Nonperturbative contributions into polarization operator have the form
(restricted by d=6 operators):
$$\Pi_{nonpert}(Q^2) = \frac{1}{(4m^2)^2} \langle 0\mid
\frac{\alpha_s}{\pi} G^2 \mid 0 \rangle [~f^{(0)}(z) +af^{(1)}
(z)~] + $$
\be
+\frac{1}{(4m^2)^3} g^3 f^{abc} \langle 0 \mid G^a_{\mu\nu}
G^b_{\nu\lambda} G^c_{\lambda \mu} \mid 0 \rangle F(z),~~
z=-\frac{Q^2}{4m^2}\label{65}\ee


\begin{figure}[tb]
\hspace{30mm} \epsfig{file=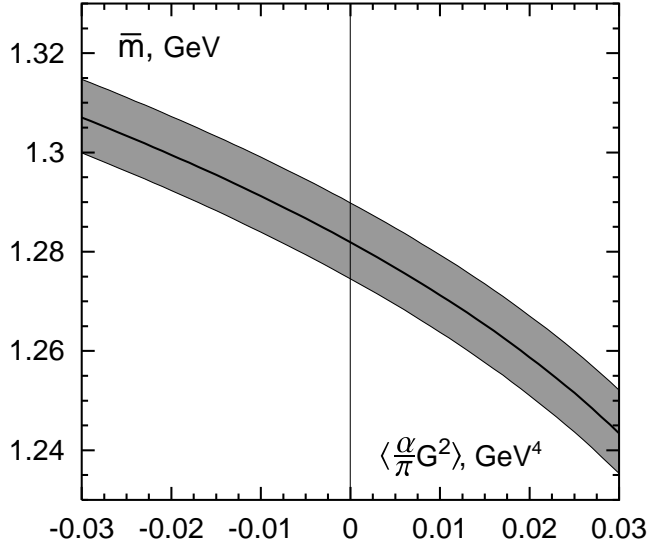, width=85mm} \caption{The
dependence of $\overline{m}(\overline{m})$ on $\langle 0 \vert
\alpha s/\pi)G^2 \vert 0 \rangle$ obtained at $n = 10$,~ $Q^2 =
0.98 \cdot 4m^2$ and $\alpha_s(Q^2 + \overline{m}^2)$. }
\end{figure}


\begin{figure}[tb]
\hspace{0mm} \epsfig{file=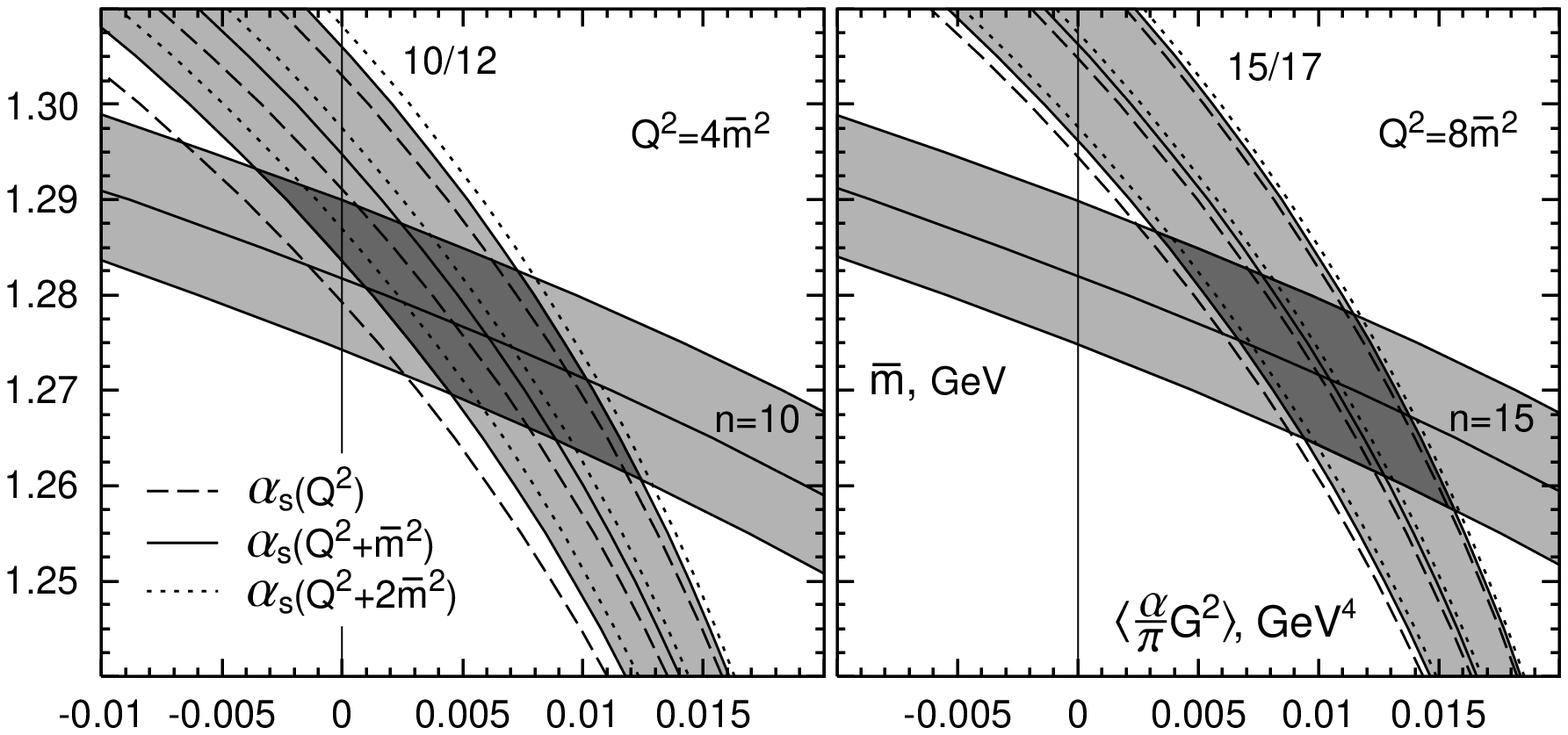, width=150mm} \caption{The
dependence of $\overline{m}(\overline{m})$ on $\langle 0 \vert
(\alpha s/\pi) G^2 \vert 0 \rangle$ obtained from the moments
(horizontal bands) and their ratios (vertical bands) at different
$\alpha_s$. The left-hand figure: $Q^2 = 4 \overline{m}^2$, ~ $n =
10$, ~ $M_{10}/M_{12}$; the right-hand figure -- $Q^2 = 8
\overline{m}^2$, ~ $n = 15$, ~ $M_{15}/M_{17}$. }
\end{figure}
\noindent
 Functions $f^{(0)}(z)$, ~ $f^{(1)}(z)$ and $F(z)$ were
calculated in [5], [54], [55], respectively. The use of the quark
pole mass is, however, inacceptable. The matter is that in this
case the PT corrections to moments are very large in the region of
interest and perturbative series  seems to diverge. Thus, for
instance, at
\be
n=10 \; , \; Q^2=4m^2 :  \qquad {M^{(1)}\over M^{(0)}}=13.836 \; ,
\qquad {M^{(2)}\over M^{(0)}}=193.33 \; , \qquad {M^{(G,1)}\over
M^{(G,0)}}=13.791 \label{66} \ee (here $M^{(k)}$ mean the
coefficients at the contributions of terms $\sim a^k$ to the
moments, $M^{(G,k)}$ - are the similar coefficients for gluonic
condensate contribution. In the region of interest $a \sim 0.1$).
At $Q^2 = 0$ the situation is even worse. So, it is reasonable to
turn to to $\overline{MS}$ mass $\overline{m}(\mu^2)$, taken at
the point $\mu^2 = \overline{m}^2$.  After turning to the
$\overline{MS}$ mass $\overline{m}(\overline{m}^2)$ we get instead
of (66):
\be
n=10 \; , \; Q^2=4{\bar m}^2:  \qquad {{\bar
M}^{(1)}\over {\bar M}^{(0)}}=0.045 \; , \qquad {{\bar M}^{(2)}\over {\bar
M}^{(0)}}=1.136 \; , \qquad {{\bar M}^{(G,1)}\over {\bar M}^{(G,0)}}=-1.673
\label{67} \ee

At $a \sim 0.1$ and at the ratios of moments given by (67) there is a good
reason to believe that the PT series well converges. Such a good convergence
holds (at $n > 5$) only in the case of large enough $Q^2$, at $Q^2 = 0$  one
does not succeed in finding such $n$, that perturbative corrections,
$\alpha_s$ corrections to gluonic condensates and the term $\sim \langle G^3
\rangle$ contribution would be simultaneously small.

It is also necessary to choose scale - normalization point $\mu^2$
where $\alpha_s(\mu^2)$ is taken. In (64)~ $R(s)$ is a physical
value and cannot depend on $\mu^2$. Since, however, we take into
account in (64) only three terms, at unsuitable choice of $\mu^2$
such $\mu^2$ dependence may arise due to neglected terms. At large
$Q^2$  the natural choice is $\mu^2 = Q^2$. It can be thought that
at $Q^2 = 0$ the reasonable scale is $\mu^2 = \overline{m}^2$,
though some numerical factor is not excluded in this equality.
That is why it is reasonable to take interpolation form
\be
\mu^2 = Q^2+\overline{m}^2,\label{68}\ee but to check the
dependence of final results on a possible factor at
$\overline{m}^2$. Equalling theoretical value of some moment at
fixed  $Q^2$ (in the region where $M^{(1)}_n$ and $M^{(2)}_n$ are
small) to its experimental value one can find the dependence of
$\overline{m}$ on $\langle(\alpha_s/\pi)G^2 \rangle$ (neglecting
the terms $\sim \langle G^3 \rangle$). Such a dependence for $n =
10$ and $Q^2/4 m^2 = 0.98$ is presented in Fig.6.

To fix both $\overline{m}$ and $\langle(\alpha_s/\pi) G^2 \rangle$
one should, except for moments, take their ratios. Fig.7 shows the
value of $\overline{m}$ obtained from the moment $M_{10}$ and the
ratio $M_{10}/M_{12}$ at $Q^2 = 4 m^2$ and from the moment
$M_{15}$ and the ratio $M_{15}/M_{17}$ at $Q^2 = 8 m^2$. The best
values of masses of charmed quark and gluonic condensate are
obtained from fig.7:
\be
{\bar m}({\bar m}^2)\,=\,1.275\pm 0.015 \, {\rm GeV} \; , \qquad
\left< {\alpha_s\over \pi} G^2\right>\,=\,0.009\pm 0.007 \, {\rm
GeV}^4 \label{69}\ee Up to now the corrections $\sim \langle G^3
\rangle$ were not taken into account. It appears that in the
region of $n$ and $Q^2$ used to find $\overline{m}$ and gluonic
condensate they are comparatively small and, practically, not
changing $\overline{m}$, increase $\langle ((\alpha_s/\pi)G^2$ by
$10-20\%$ if the term $\sim \langle G^3 \rangle$ is estimated
according to (13) at $\rho_c = 0.5 fm$.

It should be noted that improvement of the accuracy of $\Gamma_{J/\psi \to
ee}$ would make it possible to precise the value of gluonic condensate: the
widths of horizontal bands in fig.7 are determined mainly just by this
error. In particular,  this, perhaps, would allow one to exclude the zero
value of gluonic condensate, that would be extremely important.
Unfortunately, eq.(69) does not allow one to do it for sure. Diminution of
theoreticl errors which determine the width  of vertical bands seems to be
less real.

\vspace{3mm}

\begin{center}
{\bf \large 6. Conclusion}
\end{center}

In this paper I compare the results of the recent precise
measurements of $\tau$-lepton hadronic decays [24]-[26] with QCD
predictions in the low energy region. The perturbative terms up to
$\alpha^3_s$ and the terms of the operator product expansion (OPE)
up to d=8 were taken into account. It is shown that QCD with the
account of OPE terms agrees with experiment up to $\sim 2\%$ at
the values of the complex Borel parameter $\vert M^2 \vert >
0.8-1.0 GeV^2$ in the left-hand semiplane of the complex plane. It
was found:\\ 1. The values of the QCD coupling constant
$\alpha_s(m^2_{\tau}) = 0.355 \pm 0.025$ from the total
probability of $\tau$-decays and $\alpha_s(m^2_{\tau}) = 0.330$
from the sum rules at low energies. (The latter value corresponds
to $\alpha_s(m^2_z) = 0.118$).\\ 2. The value of the quark
condensate square (assuming factorization) $$ \alpha_s \langle
\mid \overline{\psi}\psi \mid 0\rangle^2 = (2.25 \pm 0.70)\cdot
10^{-4}~\mbox{GeV}^6$$  and of quark-gluon condensate of d=8.\\ 3.
The value of gluonic condensate:

a) from the $\tau$-decay data: $$\langle 0\mid
\frac{\alpha_s}{\pi} G^2 \mid 0 \rangle = (0.006\pm
0.012)~\mbox{GeV}^4$$

b) from the sum rules for charmonium
 $$\langle 0\mid
\frac{\alpha_s}{\pi} G^2 \mid 0 \rangle = (0.009\pm 0.007)~
GeV^4$$ It is shown that the sum rules for charmonium are in
agreement with experiment when accounting for perturbative
corrections $\sim \alpha^2_s$ and for OPE terms proportional to
$\langle (\alpha_s/\pi) G^2 \rangle$ and to $\langle G^3 \rangle$.

The main conclusion is that in the range of low-energy phenomena under
consideration, perturbation theory and operator expansion, i.e. the idea of
vacuum condensates in QCD is in an excellent agreement with experiment
starting from $Q^2 \sim 1 GeV^2$.

I am deeply indebted to K.N.Zyablyuk who had made the main calculations in
papers [27, 28], the results of which I used here.

The paper is supported by the grants CRDF RP2-2247, INTAS-2000-587 and RFFI
00-02-17808.

\newpage


\begin{thebibliography}{99}
\bibitem{1} P.A.Aleksandrov, Akademik Anatoly Petrovich Aleksandrov:
Pryamaya Rech (in Russian), Academician Anatoly Petrovich
Aleksandrov: Direct speech. M.Nauka, 2001, p.177.
\bibitem{2} D.Holloway, Stalin and the bomb, Yale Univ.Press, New
Haven \& London, 1994, chapt.15, sec.6.
\bibitem{3} B.L.Ioffe, A top secret assignment, Novy Mir,
1999, No.6, p.161 (in Russian). English translation in: At the
frontier of particle physics, Handbook of QCD, Boris Ioffe
Festschrift, ed. by M.Shifman, World Svientific, 2001, v.1.
\bibitem{4}
A.A.Belavin, A.M.Polyakov, A.S.Schwarz, Yu.S.Tyupkin, Phys.Lett.B,
{\bf 59}, 85 (1975).\bibitem{5} M.A.Shifman, A.I.Vainstein,
V.I.Zakharov, Nucl.Phys.B {\bf 147}, 385,448 (1979). \bibitem{6}
V.A.Novikov, M.A.Shifman, A.I.Vainstein and V.I.Zakharov,
Nucl.Phys.B {\bf 249}, 445 (1985). \bibitem{7} M.A.Shifman, {\it
Lecture at 1997 Yukawa International Seminar}, Kyoto, 1997,
Suppl.Prog.Theor.Phys., 1998, Vol.131, p.1.\bibitem{8}
M.Gell-Mann, R.J.Oakes, B.Renner, Phys.Rev. {\bf 175}, 2195
(1968). \bibitem{9} B.L.Ioffe, Usp.Fiz.Nauk {\bf 171}, 1273
(2001).\bibitem{10} S.Weinberg, in: {\it A Festschrift for
I.I.Rabi}, Trans.New York Acad.Sci., Ser.2,  Vol.38, p.185, 1977.
\bibitem{11} H.Leutwyler, Journ. Moscow Phys.Soc., {\bf 6}, 1
(1996). \bibitem{12} B.L.Ioffe, Nucl.Phys.B  {\bf 188}, 317
(1981); {\bf 192}, 591 (1982). \bibitem{13} H.Leutwyler in: {\it
At the Frontier of Particle Physics}, Handbook of QCD, Boris Ioffe
Festschrift, ed. by M.Shifman, World Scientific,  Vol.1, p.271,
2001.
\bibitem{14} U.Meissner, ibid, p.417.
\bibitem{16} P.Gerber, H.Leutwyler, Nucl.Phys.B {\bf 321}, 387
(1989).
\bibitem{15} P.Chen et al., Phys.Rev.D {\bf 64}, 014503 (2001).
\bibitem{17}
B.L.Ioffe, {\it  Lecture at St.Petersburg Winter School on
Theoretical Physics}, Febr.1998, Surveys in High Energy Physics,
Vol.14, p.89, 1999. \bibitem{18} V.M.Belyaev, B.L.Ioffe, ZhETF
{\bf 83}, 876 (1982).
\bibitem{19} V.A.Novikov, M.A.Shifman, A.I.Vainstein,
V.I.Zakharov, Phys.Lett.B {\bf 86}, 347 (1979).
\bibitem{20} B.L.Ioffe, A.V.Smilga, Nucl.Phys.B {\bf 232}, 109
(1984). \bibitem{21} V.M.Belyaev, Ya.I.Kogan, Yad.Fiz. {\bf 40},
1035 (1984).\\ I.I.Balitsky, A.V.Kolesnichenko, A.V.Yung, Yad.Fiz.
{\bf 41}, 282 (1985).
\bibitem{22} V.M.Belyaev, Ya.I.Kogan, Pis'ma v  ZhETF {\bf 37}, 611
(1983).
\bibitem{23} B.L.Ioffe, A.G.Oganesian, Phys.Rev.D {\bf 57}, R6590
(1998).\bibitem{24} ALEPH Collaboration, R.Barate et al.,
Eur.Phys.J.C {\bf 4}, 409 (1998).
\bibitem{25} OPAL Collaboration, K.Ackerstaff et al., Eur.Phys.J.C
{\bf 7}, 571 (1999); G.Abbiendi et al., ibid, {\bf 13}, 197
(2002).
\bibitem{26} CLEO Collaboration, S.J.Richichi et al., Phys.Rev.D
{\bf 60}, 112002 (1999). \bibitem{27} B.L.Ioffe, K.N.Zyablyuk,
Nucl.Phys.A {\bf 687}, 437 (2001). \bibitem{28} B.V.Geshkenbein,
B.L.Ioffe, K.N.Zyablyuk, Phys.Rev.D {\bf 64}, 093009 (2001).
\bibitem{29} A.Pich, Proc. of QCD94 Workshop, Monpellier, 1944;
Nucl.Phys.B (proc.Suppl) {\bf 39}, 396 (1995). \bibitem{30}
W.J.Marciano, A.Sirlin, Phys.Rev.Lett. {\bf 61}, 1815 (1988).
\bibitem{31} E.Braaten, Phys.Rev.Lett. {\bf 60}, 1606 (1988);
Phys.Rev.D {\bf 39}, 1458 (1989). \bibitem{32} S.Narison, A.Pich,
Phys.Lett.B {\bf 211}, 183 (1988).
\bibitem{33} F.Le Diberder, A.Pich, Phys.Lett.B {\bf 286}, 147
(1992). \bibitem{34} K.G.Chetyrkin, A.L.Kataev, F.V.Tkachov,
Phys.Lett.B {\bf 85}, 277 (1979); M.Dine, J.Sapirshtein,
Phys.Rev.Lett. {\bf 43} 668 (1979); W.Celmaster, R.Gonsalves,
ibid, {\bf 44}, 560 (1980). \bibitem{35} L.R.Surgaladze,
M.A.Samuel, Phys.Rev.Lett. {\bf 66}, 560 (1991);\\ S.G.Goryshny,
A.L.Kataev, S.A.Larin, Phys.Lett.B {\bf 259}, 144 (1991).
\bibitem{36}
A.L.Kataev, V.V.Starshenko, Mod.Phys.Lett.A {\bf 10}, 235 (1995).
\bibitem{37} O.V.Tarasov, A.A.Vladimirov, A.Yu.Zharkov,
Phys.Lett.B  {\bf 93}, 429 (1980); S.A.Larin, J.A.M.Vermaseren,
ibid, {\bf 303}, 334 (1993). \bibitem{38} T.van Ritbergen,
J.A.M.Vermaseren, S.A.Larin, Phys.Lett.B {\bf 400}, 379 (1997).
\bibitem{39} K.G.Chetyrkin, S.G.Gorishny, V.P.Spiridonov,
Phys.Lett.B {\bf 160}, 149 (1985). \bibitem{40} L.-E.Adam,
K.G.Chetyrkin, Phys.Lett.B {\bf 329}, 129 (1994). \bibitem{41}
E.Braaten, C.S.Lee, Phys.Rev.D {\bf 42}, 3888 (1990). \bibitem{42}
K.Hagiwara et al., Particle Data Groop, Phys.Rev.D {\bf 66},
010001 (2002).\bibitem{43} ALEPH Collaboration, R.Barate et al.,
Eur.Phys.J.C {\bf 11}, 599 (1999).
\bibitem{44} OPAL Collaboration, G.Abbiendi et al., Eur.Phys.J.C
{\bf 19}, 653 (2001).
\bibitem{45} I.L. Solovtsov, D.V.Shirkov, Phys.Rev.Lett. {\bf 79},
1209 (1997), Teor.Mat.Fiz. {\bf 120}, 1210 (1999).
\bibitem{ab} B.V.Geshkenbein, B.L.Ioffe, Pis'ma v ZhETF, {\bf 70},
167 (1999).
\bibitem{46}
K.G.Chetyrkin, S.Narison, V.I.Zakharov, Nucl.Phys.B {\bf 550}, 353
(1999).
\bibitem{47} T.Shafer, E.V.Shuryak, Rev.Mod.Phys. {\bf 70}, 323
(1998).
\bibitem{48} B.L.Ioffe, K.N.Zyablyuk, hep-ph/0207183.
\bibitem{49} V.B.Berestetsky, I.Ya.Pomeranchuk, JETP {\bf 29}, 864
(1955).
\bibitem{50} J.Schwinger, Particles, Sources, Fields, Addison-Wesley
Publ.,
1973, V.2. \bibitem{51} A.H.Hoang, J.H.Kuhn, T.Teubner,
Nucl.Phys.B {\bf 452}, 173 (1995);\\ K.G.Chetyrkin, J.H.Kuhn,
M.Steinhauser, Nucl.Phys.B {\bf 482}, 213 (1996);\\ K.G.Chetyrkin
et al., Nucl.Phys.B {\bf 503}, 339 (1997);\\ K.G.Chetyrkin et al.,
Eur.Phys.J.C {\bf 2}, 137 (1998).
\bibitem{52} D.J.Broadhurst et al., Phys.Lett.B {\bf 329}, 103
(1994). \bibitem{53} S.N.Nikolaev, A.V.Radyushkin, Yad.Fiz. {\bf
39}, 147 (1984).


\end{thebibliography}
\end{document}